# Four-node Relay Network with Bi-directional Traffic Employing Wireless Network Coding with Pre-cancellation

Su Kiang Kuek
National University of Singapore

Chau Yuen, Woon Hau Chin
Institute for Infocomm Research
cyuen@i2r.a-star.edu.sg, w.h.chin@ieee.org

*Abstract*—Network coding has the potential to improve the overall throughput of a network by combining different streams of data and forwarding them. In wireless networks, the wireless channel provide an excellent medium for physical layer network coding as signals from different transmitters are combined automatically by the wireless channel. In such scenarios, it would be interesting to investigate protocols and algorithms which can optimally relay information. In this paper, we look at a four-node two-way or bi-directional relay network, and propose a relay protocol which can relay information efficiently in this network.

## I. Introduction

The broadcast nature of the wireless channel facilitates the employment of network coding to enhance the throughput of a wireless network. While most network coding research have focused on bit level manipulation of the data stream, recently, there have been interests in "analog network coding" (ANC), or "physical layer network coding". ANC is particularly useful in wireless networks as the wireless channel acts as a natural implementation of network coding by summing the wireless signals over the air. In this paper, we only consider ANC, and shall use ANC and wireless network coding (WNC) synonymously.

In this paper, we consider bi-directional traffic flows where two nodes tries to send messages to each other through the relay network in between them. The simplest case, a three node (two sources/destinations, one relay) relay network was studied in [1] and [2]. Most of the available literature continue to consider the three-node bi-directional relay network. Interestingly, using wireless network coding in an effective way for more than three nodes is unknown in the literature.

In this paper, we consider a four node wireless relay network with bidirectional traffic. Unlike its three-node counter part, four node network poses a challenging task, as this setup can no longer simply receive and broadcast. Here, we propose to solve this problem by proposing a technique call "precancellation". By combining this technique with wireless network coding, we show that we are able to achieve a significant throughput gain in a four node relay network.

## II. Three Node Bi-Directional Relay Networks

We first illustrate the problem of bi-directional traffic relaying by introducing the classical three node bidirectional relay case. Each of the relays involved are assumed to be half duplex, *i.e.* each relay can only either transmit or receive at any one time. To keep the problem simple, we also assume that the transmissions are noiseless at this stage.

In this network, we have two nodes (A and C) acting as both the information source and sink. These two nodes are unable to communicate directly with each other, but are able to transmit to and receive information from the relay node B.

The most straightforward solution to the problem is for one of the party (say A) to first transmit the message, $x$, to B and B to relay the message to C. Subsequently, C will transmit $y$ to B and B will relay the message to A. This solution is illustrated in Figure 1, where the shaded boxes indicate the transmitting node. From the figure, we can observe that this solution requires four time slots to transmit two sets of information.

Alternatively, we can consider the solution in [2], which is illustrated in Figure 2. In this case, we allow both A and C to transmit simultaneously to B. Due to the superposition nature of the wireless channel, B receives $x + y$. B can then broadcast $x + y$ to both A and C at the same time. Since A has knowledge of x, it can extract y from the received $x + y$. Likewise, C is able to decode x.

By using WNC, the number time slots required for the transmission of two messages is reduced from four to two, doubling the throughput of the system.

| Time | Node A | | Node B | | Node C |
|------|--------|---|--------|---|--------|
| 1 | | x → | x | | |
| 2 | | ← x | x | x → | x |
| 3 | | | x  y | ← y | x |
| 4 | y | ← y | x  y | y → | x |

Figure 1 A bi-directional transmission for 3-node relay network without wireless network coding

| Time | Node A | | Node B | | Node C |
|---|---|---|---|---|---|
| 1 | | → x | x  y | ← y | |
| 2 | y | ← x+y | x  y | x+y → | x |

Figure 2 A bi-directional transmission for 3-node relay network with wireless network coding

### III. FOUR NODE BI-DIRECTIONAL RELAY NETWORKS

Unlike the three-node relay network described in the previous section, the four node (two sources/destinations, two relays) case is much more complicated. The relays will have to coordinate such that only one of them is transmitting at any one time, and they can no longer be passive by simply receiving and broadcasting the messages they receive.

For normal bi-directional relaying in this case, the optimal message transmission scheme without WNC would be to have both sources, A and D transmitting, $x_1$ and $y_1$ respectively, at the same time in the first time slot. In the next time slot, B will broadcast $x_1$ while C listens in and keeps $y_1$ in its buffer. Subsequently, C will transmit $y_1$, and both B and C transmit $y_1$ and $x_1$ respectively in the last time slot. This process is illustrated in Figure 3.

For WNC, the scenario is also not straightforward. The relays can no longer simply receive and broadcast the messages they received as that will result in multiple copies of messages being summed up and this makes the decoding of the message at the destination nodes becomes difficult. This problem can be observed in time slot 5 in Figure 4, where the problematic message is highlighted.

To overcome the problem, pre-cancellation at the source nodes is proposed. As shown in Figure 5, at time slot 3, instead of transmitting $x_2$ from A to B, A transmits $x_2 - x_1$ so as to cancel the previous message from the relay. Hence from time slot 3 onwards, we notice a pattern in the transmission. At the source node, you transmit $x_k - x_{k-1}$. By doing so, at node B, it will relay $x_k + y_{k-1}$; and node C, it will relay $x_k + y_k$.

From the new transmission scheme we can infer the following conjecture:

*Conjecture 1:* In a relay network, the throughput gain provided by wireless network coding with pre-cancellation is approximately double of the one without network coding.

*Proof:* In the three-node relay network, with wireless network coding have a 2 times throughput gain (1 msg/time vs 0.5 msg/time). In the case of four-node relay network, from time slot 5 onwards, it takes only two additional time slots to exchange two messages, i.e. about 1 msg/time. This is also double of the throughput for the case without network coding. This suggests that the larger the network size, as long as the time approaches infinity, and if you manage to encode and relay the messages properly, throughput gain of using network coding is about double of that without network coding.
∎

Next, we extend the case by taking channel coefficient into account, as shown in Figure 6. The channel coefficient between node A and B, B and C and C and D is denoted by $h_1$, $h_2$ and $h_3$ and are assume to be constant and known to all nodes throughout the period of transmission. It can be noted that the inclusion of channel coefficients do not affect the throughput gain provided by wireless network coding with pre-cancellation as discussed earlier in Conjecture 1.

Coding at the source node is further enhanced from time slot 3 onwards to include the channel coefficient for efficient pre-cancellation. As shown in Figure 6, from time slot 3 onwards, $h_2 h_2$ is coded at the transmitting source node A at odd time slot and $h_3 h_2$ is coded at the transmitting source node D at even time slot to allow efficient pre-cancellation. From time slot 3 onwards, the intermediate nodes B and C also remove the common $h_2$ channel coefficient from the received packet.

Finally, we complete the case by including the power normalization factor and noise into the transmission chart, as shown in Figure 7. It is assumed that the power normalization constant $\alpha$ and $\beta$ is the same for every odd and even time instance respectively and is known to all nodes. Noise is random and is added at every receiving node. Throughput is unaffected and $\alpha$, $\beta$ are included at every transmitting node and removed at the received node as it is a common factor.

Derivation of $\alpha$ and $\beta$

From time slot 3 onwards, the received packet at the intermediate node is denoted by U and V respectively.

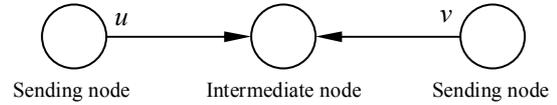

Sending node    Intermediate node    Sending node

For odd time slot $2n+1$ where $n > 0$, integer,
$$u_{odd} = \alpha h_2^2 (x_{n+1} - x_n) \\ v_{odd} = \alpha(h_1 h_2 x_n + h_3 y_n) \quad (1)$$

For even time slot $2n$ where $n > 1$, integer,
$$u_{even} = \beta h_1 h_2 (x_n + h_3 y_{n-1}) \\ v_{even} = \beta h_2 (y_n + y_{n-1}) \quad (2)$$

From $E[|u|^2 + |v|^2] = 1$, it can be shown that
$$\alpha^2 = \frac{1}{|h_2|^4 + |h_2|^2 |h_1|^2 + |h_3|^2} \\ \beta^2 = \frac{1}{|h_1|^2 |h_2|^2 + |h_3|^2 + 2|h_2|^2} \quad (3)$$

### IV. SIMULATIONS

A Matlab simulation is written to investigate the error performance of the 4 nodes two-way / bi-directional relay network with wireless network coding. It is assumed that the three channel coefficients and power normalization factors are constant and known to all nodes. White Gaussian noise is also added at every receiving node for each time instance.

Simulations were performed for 10, 50 and 100 time slots for each run for SNR from 0 to 60. It is assumed that the SNR for every node is the same.

As in Fig 8 to Fig 9, It can be observed that the bit error rate (BER) decreases as the SNR increases from 0 to 60. It can also be observed that the bit error rate graphs for packet $x$ and $y$ are not the same. This is due to the difference in the power allocated to the transmitted symbols at node A and D respectively.

From Figures 6 and 7, it could be seen that the transmission at node B introduced a double Rayleigh fading factor for the $y$ messages to node A. While the transmission of $x$ messages at node C to node D introduced a triple Rayleigh fading factor. This difference contributed to the difference in the BER of the two data streams.

The bit error rate is also lower for transmission cycle with shorter time instance. It can be seen in Figure 6 that the noise generated from each transmission starting from time slot 1 will be present and added to each subsequent transmission till the end of the cycle. Hence transmission cycle with shorter time instance will have less noise interference and result in the lower bit error rate.

The bit error rate for transmission with power normalization factor is shown in Figure 10 to 11.

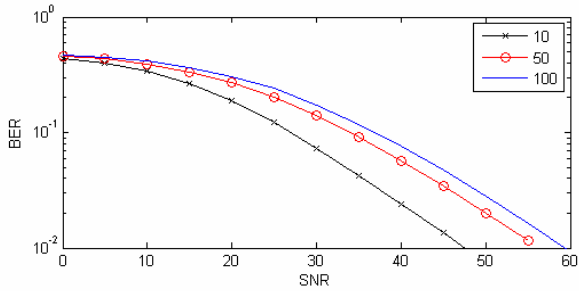
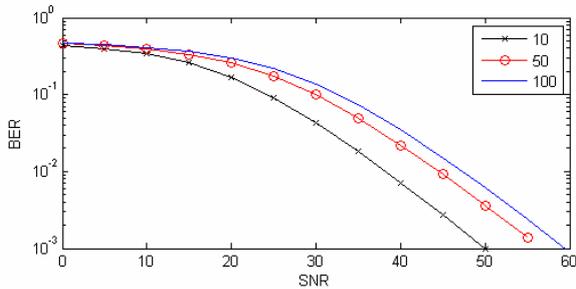

Figure 8 (upper figure) Bit error rate for packet $x$ without power normalization factor

Figure 9 (lower figure) Bit error rate for packet $y$ without power normalization factor

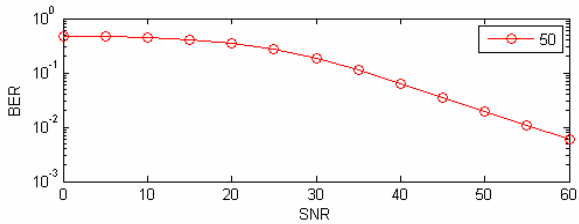
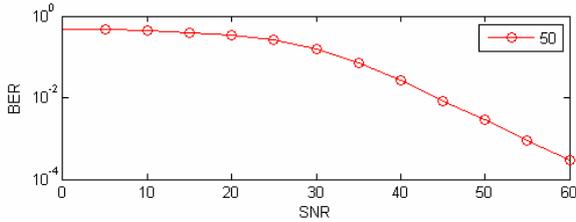

Fig 10. (upper figure) Bit error rate for packet $x$ with power normalization factor

Fig 11. (lower figure) Bit error rate for packet $y$ with power normalization factor

## V. CONCLUSION

In this paper, we proposed a protocol based on wireless network coding for a four-node relay network with two-way / bidirectional traffic. To the best of our knowledge, this is the first paper dealing with a relay network with an even number of nodes. Unlike traditional wireless network coding that simply sum up the signal at the relays, the novelty of the proposed scheme lies in the part that the source node would pre-cancel some of the prior message, such that the message would not "oscillate" within the network, and this is crucial for a four-node network. We believe this technique can be easily extended to any network that employ even number of nodes.

| Time | Node A | | Node B | | Node C | | Node D |
|---|---|---|---|---|---|---|---|
| 1 | | x1 → | x1 | | y1 | ← y1 | |
| 2 | | ← x1 | x1 | x1 → | x1   y1 | | |
| 3 | | | x1   y1 | ← y1 | x1   y1 | ← y1 | |
| 4 | y1 | ← y1 | x1   y1 | x1   y1 | x1   y1 | ← x1 | x1 |

Figure 3 A normal bi-directional 4-node relay without wireless network coding

Figure 4 A bi-directional 4-node relay with problematic message

Figure 5 A bi-directional 4-node relay with wireless network coding and pre-cancellation

Figure 6  A bi-directional 4-node relay with wireless network coding and pre-cancellation with channel coefficient taken into account

| Time | Node A | | Node B (relay) | | Node C (relay) | | Node D |
|---|---|---|---|---|---|---|---|
| 1 | | Ah1x1+N1 → | h1x1+N1/A | | | | |
| 2 | | ← Bh1(h1x1+N/A) | h1x1+N1/A | Bh2(h1x1+N/A) → | h2h1x1+h3y1+N/AB+N/B | ← Bh3y1 | |
| 3 | | Ah1h2h2(x2-x1) → | h1x1+N1/A, h1h2x2+h3y1+N/AAB+N/BA+N/A | ← Ah2(h2h1x1+h3y1+N/AB+N/B) | h2h1x1+h3y1+N/AB+N/B | Ah3(h2h1x1+h3y1+N/AB+N/B) → | x1 |
| 4 | y1 | ← Bh1(h1h2x2+h3y1+N/AAB+N/BA+N/A) | h1x1+N1/A, h1h2x2+h3y1+N/AAB+N/BA+N/A | Bh2(h1h2x2+h3y1+N/AAB+N/BA+N/A) → | h2h1x1+h3y1+N/A+N/B, h2h1x2+h3y2+N/AABB+N/BAB+N/AB+N/B | ← Bh3h2(y2-y1) | x1 |
| 5 | y1 | Ah1h2h2(x3-x2) → | h1x1+N1/A, h1h2x2+h3y1+N/AAB+N/BA+N/A, h1h2x3+h3y2+N/AABBA+N/ABA+N/BA+N/A | ← Ah2(h2h1x2+h3y2+N/AABB+N/BAB+N/AB+N/B) | h2h1x1+h3y1+N/A+N/B, h2h1x2+h3y2+N/AABB+N/BAB+N/AB+N/B | Ah3(h2h1x2+h3y2+N/AABB+N/BAB+N/AB+N/B) → | x1 x2 |
| 6 | y1 y2 | ← Bh1(h1h2x3+h3y2+N/AABBA+N/BABA+N/ABA+N/BA+N/A) | h1x1+N1/A, h1h2x2+h3y1+N/AAB+N/BA+N/A, h1h2x3+h3y2+N/AABBA+N/BABA+N/ABA+N/BA+N/A | Bh2(h1h2x3+h3y2+N/AABBA+N/BABA+N/ABA+N/BA+N/A) → | h2h1x1+h3y1+N/A+N/B, h2h1x2+h3y2+N/AABB+N/BAB+N/AB+N/B, h2h1x3+h3y3+N/AABBAB+N/BABAB+N/... | ← Bh3h2(y3-y2) | x1 x2 |

Figure 7. A bi-directional 4-node relay with wireless network coding and pre-cancellation, taking into account the channel effects and power normalization.